\documentclass[11pt]{article}
\usepackage{amsmath,amssymb,amsthm,mathtools}
\usepackage{float}   
\usepackage{hyperref}
\usepackage{cleveref}
\usepackage{geometry} 
\usepackage{changepage}
\usepackage{float}      
\usepackage{tabularx} 
\usepackage{amsmath}   
\usepackage{amssymb} 
\usepackage{xcolor}
\usepackage{booktabs} 
\usepackage{caption}
\usepackage{booktabs}
\usepackage{makecell}

\newtheorem{theorem}{Theorem}

\newcommand{\suppSone}{\href{run:Supplementary_File_S1.pdf}{Supplementary File S1}}
\newcommand{\suppStwo}{\href{run:Supplementary_File_S2.pdf}{Supplementary File S2}}
\newcommand{\suppSthree}{\href{run:Supplementary_File_S3.pdf}{Supplementary File S3}}
\newcommand{\suppSfour}{\href{run:Supplementary_File_S4.pdf}{Supplementary File S4}}
\newcommand{\suppSfive}{\href{run:Supplementary_File_S5.pdf}{Supplementary File S5}}
\newcommand{\suppSfiveRaw}{\href{run:Supplementary_File_S5_Synchronous_STG_Feature_Matrix_89x34.csv}{synchronous raw 89x34 CSV matrix}}
\newcommand{\suppSfiveRetained}{\href{run:Supplementary_File_S5_Synchronous_Retained_Feature_Matrix_89x20_raw_scale.csv}{synchronous retained 89x20 raw-scale CSV matrix}}
\newcommand{\suppSfiveScaled}{\href{run:Supplementary_File_S5_Synchronous_RobustScaled_Feature_Matrix_89x20.csv}{synchronous robust-scaled 89x20 CSV matrix}}
\newcommand{\suppSfiveClusters}{\href{run:Supplementary_File_S5_Synchronous_Cluster_Assignments.csv}{synchronous cluster-assignment CSV}}
\newcommand{\suppSfiveEmbeddings}{\href{run:Supplementary_File_S5_Synchronous_Embedding_Coordinates.csv}{synchronous embedding-coordinate CSV}}
\newcommand{\suppSsix}{\href{run:Supplementary_File_S6.pdf}{Supplementary File S6}}
\newcommand{\suppSsixRaw}{\href{run:Supplementary_File_S6_Asynchronous_STG_Feature_Matrix_89x55.csv}{asynchronous raw 89x55 CSV matrix}}
\newcommand{\suppSsixScaled}{\href{run:Supplementary_File_S6_Asynchronous_RobustScaled_Feature_Matrix_89x39.csv}{asynchronous robust-scaled 89x39 CSV matrix}}
\newcommand{\suppSsixClusters}{\href{run:Supplementary_File_S6_Asynchronous_Cluster_Assignments.csv}{asynchronous cluster-assignment CSV}}
\newcommand{\suppSsixEmbeddings}{\href{run:Supplementary_File_S6_Asynchronous_Embedding_Coordinates.csv}{asynchronous embedding-coordinate CSV}}

\title{Reducing Boolean Networks via Analysis of Dynamic Network Subgraph Behavior

}
\author{
Soodabeh Zakeri\\
{\small Iranian Society of Clinical Oncology}\\
{\small Tehran, Iran}\\
{\small \texttt{soodabe.zakeri@gmail.com}}
\and
Mohieddin Jafari\\
{\small Department of Pharmacology}\\
{\small Faculty of Medicine, University of Helsinki}\\
{\small Helsinki, Finland}\\
{\small \texttt{mohieddin.jafari@helsinki.fi}}
}
\date{\today}

\begin{document}

\maketitle  

\begin{abstract}
Boolean networks provide a compact framework for modeling regulatory systems, yet their rapidly expanding state spaces make systematic dynamical analysis challenging. Here, we systematically enumerate all non-isomorphic two-node signed regulatory subgraphs with their admissible Boolean update rules and exhaustively characterize their state-transition graphs (STGs) under both synchronous and asynchronous updating. By extracting quantitative dynamical descriptors, we move beyond static wiring diagrams toward a behavior-based classification of network structures. The synchronous analysis identifies deterministic fixed-point and cyclic attractor regimes, while the asynchronous analysis captures one-node-at-a-time dynamics through terminal strongly connected components of the asynchronous STGs. Comparing the two update schemes distinguishes dynamical behaviors that are robust across update assumptions from those that depend on synchronous updating. Despite the diversity of graph-rule combinations, many realizations converge to a limited repertoire of dynamical classes, revealing substantial redundancy between structural and dynamical representations. We further propose a testable framework for investigating whether such local dynamical classes can be used to characterize selected subnetworks within larger Boolean models. To facilitate exploration and reproducibility, we developed \textbf{BORNA}, an interactive web application for exploring the complete network catalogue, STGs, and synchronous and asynchronous simulations, available at \url{https://jafarilab.github.io/BORNA/}. Together, these results provide a systematic reference for minimal Boolean network dynamics and a foundation for extending behavior-based analysis to larger regulatory systems.

\end{abstract}
\section{Introduction}

Cellular function is sustained through the coordinated regulation of thousands of genes and proteins; yet, despite this complexity, many molecular elements can often be approximated as binary “on/off” switches. Building on this abstraction, Boolean networks (BNs), that, first introduced by Kauffman in the late 1960s \cite{Kauffman1969,Bloomingdale2018}, have emerged as well-studied discrete models of biological networks such as gene regulatory networks. In this framework, each gene is represented as a Boolean variable $x_i \in \{0,1\}$, where the value $1$ denotes an active state and $0$ denotes an inactive state. The state of each variable at time $t+1$ is determined by a Boolean function $f^{(i)}$ of the states of its regulators at time $t$. Formally, if there are edges from nodes $x_{j_1}, x_{j_2}, \ldots, x_{j_l}$ to $x_i$, then the update rule is written as 
\begin{equation}
x_i(t+1) = f^{(i)}(x_{j_1}(t), x_{j_2}(t), \ldots, x_{j_l}(t)), \quad i = 1, \ldots, n.
\end{equation} 
Thus, a BN can be viewed as a directed graph in which nodes represent genes and edges capture regulatory interactions. Despite their simplicity, BN models have become a cornerstone in systems biology, providing a scalable and interpretable framework to study the qualitative dynamics of complex regulatory systems \cite{Tran2016,Devloo2003}.

BN models provide a powerful abstraction for representing regulatory and signaling processes in biology, particularly when quantitative kinetic parameters are incomplete or unavailable. Their simplicity allows the modeling of large gene regulatory and signaling systems by focusing on the combinatorial logic of interactions rather than precise reaction rates \cite{saadatpour2013boolean,Jafari2017CentralDogma}. \\However, this abstraction comes with a significant computational drawback: the size of the state space grows exponentially with the number of variables. A network of $n$ nodes yields $2^n$ possible system states, and biologically realistic networks often include dozens to hundreds of nodes. Consequently, exhaustive exploration of the state space quickly becomes infeasible, and direct identification of attractors, which are the long-term behaviors corresponding to biologically meaningful phenomena such as cell fates, stable signaling responses, or oscillatory cycles, is computationally prohibitive \cite{Saadatpour2013, Naldi2011,DubrovaTeslenko2011}. This ``state space explosion'' problem limits the practical applicability of Boolean modeling unless strategies are introduced to reduce model complexity. Without reduction, the analysis of large-scale networks becomes either computationally intractable or reliant on partial sampling, which may miss attractors of biological importance. Furthermore, this computational burden hampers model validation against experimental data, parameter exploration, and the integration of network dynamics into larger systems-level analyses \cite{VelizCuba2011Reduction,Flottmann2013,Huang2009}. 

Reduction methods have been developed to address the inherent complexity of BN models by simplifying network structures and logical rules while preserving their essential dynamical properties\cite{Melkman2010,Milo2002}. These approaches exploit features such as nodes that stabilize independently of initial conditions and the algebraic simplification of logical functions, enabling variable reduction while preserving steady states or, in some cases, attractor structure \cite{VelizCuba2014SteadyStates, Saadatpour2013,Klarner2017}. From a bioinformatics perspective, such strategies are crucial for scaling BN analysis to biologically realistic models, where exact or computationally tractable approximations of attractors would otherwise be infeasible \cite{mendoza2006network, samaga2010logic,Li2004,chaouiya2025model}. In addition, reduced models facilitate comparison across qualitative and quantitative frameworks (Boolean, multivalued logical, or continuous) and produce interpretable cores that highlight the regulatory circuits driving biological outcomes. Consequently, reduction techniques serve not only as computational tools but also as a bridge between qualitative modeling and biologically meaningful insight\cite{Wang2012Overview,Tamura2009,Thomas1990,VelizCuba2013ANDNOT}.

Several representative approaches illustrate these principles. Saadatpour et al. developed a Boolean network reduction method that conserves fixed points and complex attractors of general asynchronous Boolean models, allowing long-term dynamics of a large network to be inferred from a reduced model \cite{Saadatpour2013}. Veliz-Cuba introduced a reduction method that first simplifies Boolean functions and wiring diagrams by removing non-functional variables and edges. Subsequently, nodes without self-loops—whose update functions do not depend on their own state—are eliminated, with their regulatory influence rewired directly between regulators and targets. This procedure reduces network size while preserving attractors and other key dynamical properties \cite{VelizCuba2011Reduction}.

 This approach complements earlier strategies based on the elimination of stable variables and leaf nodes, providing a systematic framework for scaling Boolean models to larger systems\cite{Saadatpour2013}.

Complementary exact approaches include computational-algebra methods for steady-state computation \cite{VelizCuba2014SteadyStates} and SAT-based algorithms for attractor detection \cite{DubrovaTeslenko2011}. In addition, dynamically consistent reduction of logical regulatory graphs has been developed to preserve essential dynamical properties during model reduction \cite{Naldi2011}. Beyond reduction itself, Saadatpour and Albert \cite{Saadatpour2016} compared qualitative and quantitative dynamic models, showing that Boolean fixed points can correspond to stable states in continuous and piecewise affine models, while other attractors may differ across modeling formalisms.

Despite substantial progress, existing reduction methods remain largely confined to simplifying network structures and logical rules, which, while effective for attractor preservation, often overlook the rich dynamical information encoded in STGs. An STG represents the full dynamical landscape of a BN, where each node corresponds to a possible system state and directed edges describe transitions under the update scheme. This formalism captures not only steady states but also transient trajectories, feedback-driven oscillations, and basin-specific behaviors that are essential for understanding biological decision-making \cite{Saadatpour2013, Naldi2011}.
However, most reduction techniques operate directly at the BN level, providing computational tractability but offering limited insight into how such dynamical features emerge across ensembles of regulatory subgraphs \cite{VelizCuba2014SteadyStates}. In particular, repetitive or structurally similar subgraphs often recur across regulatory networks, and uncovering their shared dynamical signatures within the STG can significantly accelerate reduction by collapsing equivalent behaviors into simplified representations. Addressing this gap requires methods that integrate structural reduction with STG-based analysis, enabling the preservation of both static attractors and the broader spectrum of dynamical and subgraph-level signatures. Our work responds to this need by advancing reduction strategies that explicitly incorporate STG features and clustering, bridging the divide between qualitative abstraction and data-driven characterization of network dynamics\cite{VelizCuba2013DimensionReduction,Zanudo2013,ZanudoAlbert2013}.

In this study, we respond to the need for scalable reduction strategies by integrating structural enumeration with dynamical classification to advance the understanding of regulatory networks. Using two-node BNs as a foundational network subgraph, we systematically explore the complete set of structurally unique (non-isomorphic) signed directed graphs with self-loops, where edges represent activating or inhibitory interactions. Our framework explicitly incorporates STG features and subgraph clustering, allowing us to move beyond static wiring diagrams toward a quantitative, behavior-based classification of subgraphs. Because biological regulatory components do not necessarily update simultaneously, we analyze each enumerated Boolean Network under both deterministic synchronous updating and BoolNet-style asynchronous updating, in which one selected node is updated at a time from the current state. This comparison tests whether the inferred behavioral reduction is robust to the assumed update scheme. This approach reveals how repetitive or structurally similar subgraphs can be reduced into functional dynamic archetypes, bridging the gap between qualitative abstraction and data-driven characterization of network dynamics.

\section{Method}

\subsection{Enumeration of Non-Isomorphic Signed Directed Graphs with Loops}

This section details the combinatorial framework used to enumerate non-isomorphic signed directed graphs with loops using the classical Cauchy-Frobenius (Burnside) lemma. Our approach follows standard group-action enumeration methods \cite{vanLintWilson,HararyPalmer1973,PolyaRead1987}, avoiding cycle index polynomials while achieving closed-form, efficiently computable results.

\subsubsection{Signed Digraph Model}

Let $V = \{1, 2, \dots, n\}$ be an ordered vertex set. A signed directed graph (or \emph{signed digraph}) with loops is defined as a mapping:
\begin{equation}
A : V \times V \rightarrow \{+1, -1, 0\},
\end{equation}
where $A(i,j) = +1$ denotes activation from $i$ to $j$, $-1$ denotes inhibition, and $0$ indicates no arc. Loops are allowed, i.e., $A(i, i)$ may be nonzero.

The total number of labeled configurations is
\begin{equation}
|X| = 3^{n^2}.
\label{eq:total_config}
\end{equation}

In many applications, including the present work, one is interested only in \emph{connected} signed digraphs, meaning that at least one ordered pair $(i,j)$ with $i\neq j$ satisfies $A(i,j)\neq 0$. For $n=2$ this excludes configurations with no edge between the two vertices.

\subsubsection{Group Action and Isomorphism}
Two such directed graphs are isomorphic if one can be obtained from the other by simultaneously permuting the rows and columns of its adjacency matrix according to a permutation $\sigma \in S_n$, where $S_n$ denotes the symmetric group on $n$ elements. This induces a natural group action of $S_n$ on $X$.

\subsubsection{Burnside's Lemma Framework}

We apply Burnside’s Lemma (also known as the Cauchy–Frobenius lemma) to count orbits (i.e., isomorphism classes).

\begin{theorem}[Cauchy–Frobenius–Burnside]
Let $G$ be a finite group acting on a finite set $X$. Then the number of orbits is given by
\begin{equation}
|X/G| = \frac{1}{|G|} \sum_{g \in G} \mathrm{fix}(g),
\label{eq:burnside}
\end{equation}
where $\mathrm{fix}(g) = |\{x \in X \mid g \cdot x = x\}|$.
\end{theorem}

In our context, $G = S_n$ and $X = \{+1,-1,0\}^{n \times n}$. The challenge lies in computing $\mathrm{fix}(\sigma)$ for all $\sigma \in S_n$.

\subsubsection{Action on Ordered Pairs}

Let $\sigma \in S_n$ with cycle type $(\ell_1, \ell_2, \dots, \ell_r)$, where $\sum_i \ell_i = n$. The action of $\sigma$ on the set of ordered pairs $V \times V$ leads to orbits of matrix entries under simultaneous permutation of indices. The number of orbits under this induced action is

\begin{equation}
c_2(\sigma) = \sum_{i=1}^r \sum_{j=1}^r \gcd(\ell_i, \ell_j).
\label{eq:c2}
\end{equation}

Each orbit must be monochromatic (assigned the same sign or zero) under a configuration fixed by $\sigma$. Hence, on the \emph{full} configuration space $X$ we have
\begin{equation}
\mathrm{fix}(\sigma) = 3^{c_2(\sigma)}.
\label{eq:fix}
\end{equation}

\subsubsection{Enumeration Formula}

Substituting \eqref{eq:fix} into \eqref{eq:burnside}, we obtain the total number of non-isomorphic signed digraphs with loops as
\begin{equation}
S_n := \frac{1}{n!} \sum_{\sigma \in S_n} 3^{c_2(\sigma)},
\label{eq:final_enum}
\end{equation}
where $c_2(\sigma)$ is defined in \eqref{eq:c2}. This sequence grows asymptotically like $\Theta\left(\frac{3^{n^2}}{n!}\right)$, and for $n=2$ it gives $S_2=45$ non-isomorphic signed digraphs when \emph{all} configurations in $X$ are allowed.

In this work we further restrict attention to signed digraphs that contain at least one edge between distinct vertices. Let
\begin{equation}
X' = \{A \in X : \exists\, i\neq j \text{ with } A(i,j)\neq 0\}
\end{equation}
denote this subset, and let $S_n'$ be the number of isomorphism classes in $X'$.
\newpage
For $n=2$, the group $S_2$ consists of two permutations: the identity
$\mathrm{id}$ and the transposition $\tau=(1\,2)$. On the full space
$X$, the identity fixes all $3^4$ configurations. The transposition
$\tau$ fixes exactly those configurations for which the two diagonal
entries are equal and the two off-diagonal entries are equal; hence
there are two independent entry-orbits. Therefore,
\[
\mathrm{fix}(\mathrm{id}) = 3^4 = 81, \qquad
\mathrm{fix}(\tau) = 3^2 = 9,
\]
leading to $S_2 = \tfrac12(81+9)=45$ as above.

\begin{figure}[H]
    \centering
    \includegraphics[width=0.8\linewidth]{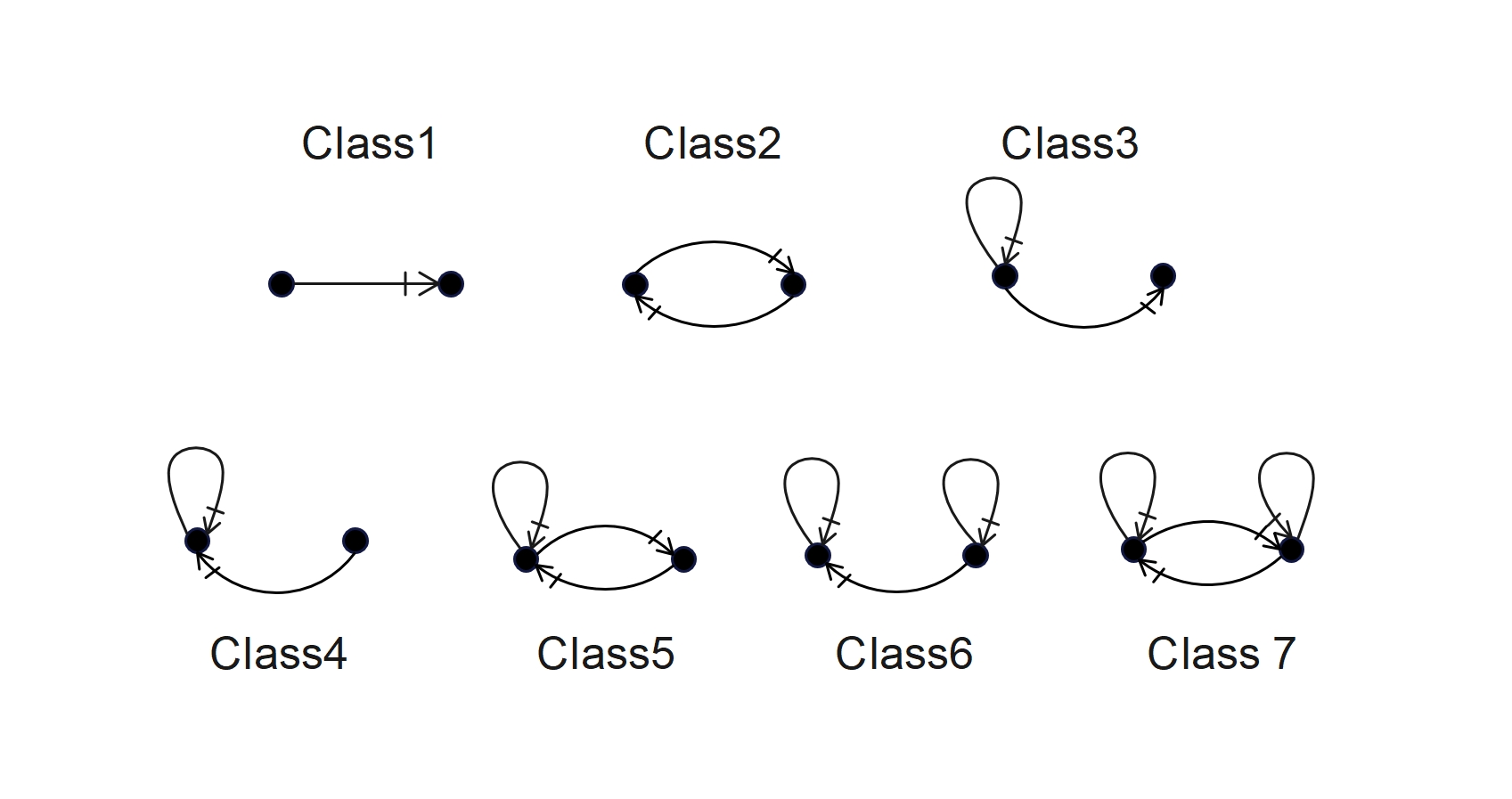}
\caption{\textbf{Structural connectivity classes of two-node regulatory subgraphs.} The schematic depicts the seven fundamental, non-isomorphic topological skeletons for connected directed graphs with $n=2$ vertices and permissible self-loops. By assigning functional signs (activation or inhibition) to each directed edge and loop, these seven base structures generate the exact 39 non-isomorphic signed digraphs evaluated in our enumeration. The connectivity types are organized by their specific combination of cross-edges and self-loops: 
    (\textbf{Class 1}) \textbf{A single directed edge} asymmetric regulation representing simple asymmetric regulation (a one-way driver-target interaction); 
    (\textbf{Class 2})\textbf{Two directed edges } forming a length-two cycle, representing bidirectional regulation or mutual feedback; 
    (\textbf{Class 3}) \textbf{Asymmetric regulation} coupled with a self-loop on the \textit{source} node (an auto-regulated node driving a downstream target); 
    (\textbf{Class 4}) \textbf{Asymmetric regulation} coupled with a self-loop on the \textit{target} node (a source node driving an auto-regulated downstream target); 
    (\textbf{Class 5}) \textbf{Mutual feedback} where exactly one of the two nodes possesses an auto-regulatory self-loop; 
    (\textbf{Class 6}) \textbf{Asymmetric regulation} connecting two nodes that \textit{both} possess auto-regulatory self-loops; and 
    (\textbf{Class 7}) \textbf{The fully connected two-node topology}, featuring bidirectional mutual feedback and self-loops on both nodes. 
    These distinct structural templates dictate the automorphism group sizes used in Burnside's Lemma and form the combinatorial basis for the subsequent Boolean rule assignments and STG extraction.}
    \label{fig:two-node-classes-connected}
\end{figure}

To obtain this, $S_2'$ we need only subtract those configurations with no cross-edge, i.e., with $A(1,2)=A(2,1)=0$:
\begin{itemize}
  \item For $\mathrm{id}$, there are $3^4=81$ total configurations, and $3^2=9$ with both off-diagonal entries equal to zero. Thus
  \[
    \mathrm{fix}_{X'}(\mathrm{id}) = 81 - 9 = 72.
  \]
  \item For $(1\,2)$, any fixed configuration must satisfy
  \[
  A(1,1)=A(2,2), \quad A(1,2)=A(2,1)=y.
  \]
  There are three choices for the diagonal value and two choices for $y\in\{+1,-1\}$ (to ensure at least one connection), hence
  \begin{equation}
    \mathrm{fix}_{X'}((1\,2)) = 3 \times 2 = 6.
  \end{equation}
\end{itemize}
Applying Burnside’s lemma on $X'$ gives
\begin{equation}
S_2' = \frac{1}{2}\bigl(\mathrm{fix}_{X'}(\mathrm{id})
       + \mathrm{fix}_{X'}((1\,2))\bigr)
     = \frac{72+6}{2}
     = 39.
\end{equation}
Thus, when we require at least one connection between the two vertices, there are
\emph{39} non-isomorphic signed digraphs with loops for $n=2$.  
These 39 graphs fall into seven underlying connectivity types, which are
depicted in Fig.~\ref{fig:two-node-classes-connected} and used below
for the Boolean rule–counting analysis.
This method aligns closely with group-action techniques used for I-graph enumeration \cite{PetkovsekZakrajsek2009} and extends the fixed-point strategy used in earlier studies on signed graphs and automorphism groups.

\subsection{Enumeration of Boolean Update Rules for Two-Node Signed Digraphs}
\label{sec:boolean-rules}

In addition to counting non-isomorphic signed digraphs, one often needs
to know how many distinct Boolean update-rule assignments each two-node
structure admits (up to relabeling).  Our approach follows the
group–action framework developed for graphical enumeration in
Section 1.6 and the standard
texts~\cite{vanLintWilson,HararyPalmer1973,PolyaRead1987}.  We focus
here on the seven connected two-node classes (those with at least one
cross-edge) shown in Fig.~\ref{fig:two-node-classes-connected}.

Write the two vertices as $A,B$, and let $k_A,k_B\in\{0,1,2\}$ be their
(signed) indegrees (a self-loop counts as one input, as does any edge
from the other node).  Define~\cite{Kauffman1969}
\[
  f_k =
  \begin{cases}
    2, & k = 0 \quad(\text{constant “always 0” or “always 1”}),\\
    1, & k = 1 \quad(\text{copy/invert the single input}),\\
    2, & k = 2 \quad(\text{AND or OR of the two inputs}).
  \end{cases}
\]
An assignment of one rule to each node is an element of a set~$X$ of
size
\begin{equation}
  |X| = f_{k_A}\,f_{k_B}.
\end{equation}

For a given two-node signed digraph $G$, the only possible non-identity
automorphism is the swap $\tau=(A\,B)$, which exists exactly when the
signed-edge pattern is symmetric under exchanging $A$ and $B$.  
The automorphism group acts on $X$ by permuting the components of the
pair $(r_A,r_B)$, and Burnside’s lemma gives
\cite{Stanley2012,cpAlgorithmsBurnside}
\begin{equation}
  \bigl|X / \operatorname{Aut}(G)\bigr|
  = 
  \frac{1}{\lvert \operatorname{Aut}(G)\rvert}
  \sum_{\sigma\in\operatorname{Aut}(G)}
    \bigl\lvert \operatorname{Fix}(\sigma)\bigr\rvert
  =
  \frac{1}{\lvert \operatorname{Aut}(G)\rvert}
  \Bigl(f_{k_A}f_{k_B}
     + [\,k_A=k_B\,]\,f_{k_A}\Bigr)
\end{equation}
Here
\begin{equation}
  \operatorname{Fix}(\mathrm{id})=f_{k_A}f_{k_B},
  \quad
  \operatorname{Fix}(\tau)=
  \begin{cases}
    f_{k_A}, & k_A=k_B,\\
    0,       & k_A\neq k_B,
  \end{cases}
\end{equation}
and $\lvert \operatorname{Aut}(G)\rvert\in\{1,2\}$.

Evaluating this expression for each of the \emph{seven} connected
two-node classes (Fig.~\ref{fig:two-node-classes-connected}) yields:

\begin{table}[H]
\centering
\begin{tabular}{c|l|c|c|c}
Class & Graph pattern (up to isomorphism) & $(k_A,k_B)$ &
$\lvert \operatorname{Aut}(G)\rvert$ & \# Boolean rules \\ \hline
1  & single cross-edge $A \to B$                      & (0,1)          & 1 & 2 \\
2  & mutual cross-edges $A \leftrightarrow B$         & (1,1)          & 2 & 1 \\
3  & self-loop on $A$ + cross-edge $A \to B$          & (1,1)          & 1 & 1 \\
4  & self-loop on $A$ + cross-edge $B \to A$          & (2,0)          & 1 & 4 \\
5  & self-loop on $A$ + mutual cross-edges            & (2,1)          & 1 & 2 \\
6  & self-loops on both nodes + single cross-edge     & (1,2)          & 1 & 2 \\
7  & self-loops on both nodes + mutual cross-edges    & (2,2)          & 2 & 3 \\
\end{tabular}
\caption{Boolean update-rule assignments per connected two-node signed
digraph class, obtained via Burnside’s lemma and the rule catalogue
$f_0=2$, $f_1=1$, $f_2=2$.}
\label{tab:boolean-two-node-connected}
\end{table}

Thus, for the seven connected structural classes in
Fig.~\ref{fig:two-node-classes-connected}, one recovers the sequence
\[
  [2,\,1,\,1,\,4,\,2,\,2,\,3]
\]
of distinct Boolean rule–assignment counts.

\subsection{General Enumeration of Boolean Update-Rule Assignments}
\label{sec:general-burnside}

Let \(G\) be a signed directed graph on \(n\) vertices labeled
\(\{1,2,\dots,n\}\).  Denote by
\[
k_i = \bigl|\{\text{incoming edges to vertex }i\}\bigr|
\]
its (signed) indegree, counting self-loops and cross-edges equally as “one input.”  
Fix a catalogue of allowed Boolean functions on \(k\) inputs and write
\[
f_k = \bigl|\{\text{allowed Boolean rules on }k\text{ inputs}\}\bigr|.
\]
For example, in the two-node case we used \(f_0=2\), \(f_1=1\), \(f_2=2\).

\medskip
Define
\[
X = \{(r_1,\dots,r_n)\mid r_i\text{ is a rule on }k_i\},
\]
so that \(\lvert X\rvert = \prod_{i=1}^n f_{k_i}\).

Let
\[
\operatorname{Aut}(G)\subseteq S_n
\]
be the automorphism group of \(G\), i.e.\ all permutations \(\sigma\)
of \(\{1,\dots,n\}\) that preserve every directed edge and its sign.

\medskip
By Burnside’s Lemma, the number of distinct Boolean-rule assignments
up to graph symmetry is
\begin{equation}
\boxed{%
\bigl|X/\operatorname{Aut}(G)\bigr|
=\frac{1}{\lvert\operatorname{Aut}(G)\rvert}
\sum_{\sigma\in\operatorname{Aut}(G)}
\bigl|\mathrm{Fix}(\sigma)\bigr|
}
\end{equation}
where \(\mathrm{Fix}(\sigma)=\{\,x\in X:\sigma\cdot x = x\}\).

\medskip
To compute \(\bigl|\mathrm{Fix}(\sigma)\bigr|\), write \(\sigma\) as a
product of disjoint cycles on \(\{1,\dots,n\}\):
\(\sigma = c_1\,c_2\cdots c_m\).  Then \(x=(r_1,\dots,r_n)\) is fixed
iff all \(r_i\) in each cycle \(c_j\) coincide.  If the common indegree
on cycle \(c_j\) is \(k_{c_j}\), there are \(f_{k_{c_j}}\) choices of
rule for that cycle, so
\begin{equation}
\bigl|\mathrm{Fix}(\sigma)\bigr|
=\prod_{j=1}^m
\begin{cases}
  f_{k_{c_j}}, & \text{if all }k_i\;(i\in c_j)\text{ coincide}\\
  0,           & \text{otherwise}
\end{cases}
\end{equation}

\paragraph{Summary of steps}
\begin{enumerate}
  \item Compute each vertex indegree \(k_i\).  
  \item Choose the local‐rule counts \(f_k\).  
  \item Determine \(\operatorname{Aut}(G)\).  
  \item For each \(\sigma\in\operatorname{Aut}(G)\):  
    \begin{itemize}
      \item Decompose \(\sigma\) into cycles \(c_j\).
      \item If all \(k_i\) in \(c_j\) agree, include factor \(f_{k_{c_j}}\); otherwise that \(\sigma\) contributes 0.
    \end{itemize}
  \item Sum these \(\lvert\mathrm{Fix}(\sigma)\rvert\) and divide by \(\lvert\operatorname{Aut}(G)\rvert\).
\end{enumerate}

This yields the exact number of Boolean update-rule assignments on \(G\),
modulo any vertex relabeling that preserves its signed‐digraph structure.

\subsubsection{Why Two Different Group Actions?  Burnside’s Lemma Applied at Two Levels}
\label{sec:two-burnsides}

We solve \emph{two} orbit–counting problems, each with its own natural
group action:

\begin{enumerate}
\item[\textbf{(A)}] \textbf{Structural enumeration.}  
      How many \emph{non-isomorphic} signed digraphs exist on $n$
      labeled vertices (in our applications, often additionally
      restricted to be connected)?
\item[\textbf{(B)}] \textbf{Rule-assignment enumeration.}  
      For a \emph{fixed} signed digraph $G$, how many distinct Boolean
      rule assignments does it admit, up to relabeling that preserves
      $G$?
\end{enumerate}

\paragraph{Problem~(A): full symmetric action.}
Let $X=\{+1,-1,0\}^{\,n\times n}$ be the set of all labeled adjacency
matrices (or, when desired, its connected subset $X'$).  
The symmetric group $S_n$ acts on $X$ by simultaneously permuting rows
and columns, so Burnside’s lemma gives
\begin{equation}
  \bigl|X/S_n\bigr|
  \;=\;
  \frac{1}{n!}\sum_{\sigma\in S_n}
  \bigl\lvert \operatorname{Fix}(\sigma)\bigr\rvert
\end{equation}
For $n=2$ this yields
$S_2=45$ isomorphism graphs on the full space, and $S_2'=39$ when we
restrict to graphs with at least one connection between the two nodes.

\paragraph{Problem~(B): automorphism action.}
Fix one graph $G$ and let $Y$ be the set of all Boolean rule
assignments.  
Two assignments are equivalent if a symmetry of $G$ carries one to the
other, i.e.\ if they lie in the same orbit of
\(
   \operatorname{Aut}(G)\subseteq S_n
\)
Hence
\begin{equation}
  \bigl|Y/\operatorname{Aut}(G)\bigr|
  \;=\;
  \frac{1}{\lvert\operatorname{Aut}(G)\rvert}
  \sum_{\sigma\in\operatorname{Aut}(G)}
  \bigl\lvert \operatorname{Fix}(\sigma)\bigr\rvert
\end{equation}
If $G$ is asymmetric then $\operatorname{Aut}(G)=\{e\}$; if $G$ is
highly symmetric the denominator reduces the count accordingly (see
Section~\ref{sec:boolean-rules}).

\paragraph{Analogy: seating a square table.}
A square table (four chairs) illustrates the difference:

\begin{itemize}
\item \emph{Structural question} (Problem A):  
      “How many shapes of empty tables?”  Only one-the square-found
      using the full $S_4$ of $24$ labelings.
\item \emph{Decorative question} (Problem B):  
      “How many colourings of the chairs are distinct up to rotating the
      table?”  Now only the four rotations matter, i.e.\
      $\operatorname{Aut}(\text{square})$.
\end{itemize}

\paragraph{Key takeaway.}
Use the big group $S_n$ to count graph structures (possibly with an
additional constraint such as connectivity); then, for each structure,
use its smaller automorphism group $\operatorname{Aut}(G)$ to
count patterns \emph{on} that structure.

\newpage

\subsection{Statistical Methods and Feature Extraction}
\label{subsec:statistical_methods}
\paragraph{From combinatorial enumeration to dynamical analysis.}
In this study, we respond to the need for scalable reduction strategies by integrating structural enumeration with dynamical classification to advance the understanding of regulatory networks. Using two-node BNs as a foundational network subgraph, we systematically explore the complete set of structurally unique (non-isomorphic) signed directed graphs with self-loops, where edges represent activating or inhibitory interactions. The preceding sections established the combinatorial foundations of this framework by enumerating (i) non--isomorphic signed digraph structures and (ii) the admissible Boolean update rule assignments associated with each structure under graph symmetries.
Consequently, every enumerated configuration can be represented dynamically by its corresponding state transition graph (STG). Having obtained the complete catalogue of such configurations for the two-node case, the next stage of the analysis shifts from combinatorial counting to dynamical characterization. In particular, we construct both synchronous and asynchronous STGs for each graph--rule pair and extract quantitative descriptors of their behavior---including attractor structure, basin organization, and transient dynamics. These dynamical features are then assembled into numerical feature matrices that form the basis for statistical and machine--learning analyses, including subgraph clustering. By incorporating STG-derived features into the analysis, our framework moves beyond static wiring diagrams toward a quantitative, behavior-based classification of subgraphs, revealing how repetitive or structurally similar motifs can be reduced into functional dynamic archetypes.
\paragraph{Scope and Update-Rule Construction.}
To systematically analyze the dynamics of the enumerated signed two-node regulatory subgraphs on \(V=\{A,B\}\), we instantiated, for each admissible signed wiring diagram and each node-wise aggregation choice, the corresponding Boolean dynamical system. Each node \(v\in V\) was assigned an update aggregation operator \(\mathsf{op}_v\in\{\mathrm{AND},\mathrm{OR}\}\). This yields four possible operator pairs \((\mathsf{op}_A,\mathsf{op}_B)\): \((\mathrm{AND},\mathrm{AND})\), \((\mathrm{AND},\mathrm{OR})\), \((\mathrm{OR},\mathrm{AND})\), and \((\mathrm{OR},\mathrm{OR})\), abbreviated as \(\mathrm{AA}\), \(\mathrm{AO}\), \(\mathrm{OA}\), and \(\mathrm{OO}\), respectively.

For a Boolean state \(x=(a,b)\in\{0,1\}^2\), with \(x_A=a\) and \(x_B=b\), the synchronous update map is
\begin{equation}
F(x)=F(a,b)=\bigl(f_A(a,b),\,f_B(a,b)\bigr)
\end{equation}
Let \(\mathcal{I}_v^+\subseteq V\) and \(\mathcal{I}_v^-\subseteq V\) denote the sets of positive and negative regulators of node \(v\), respectively. Then \(f_v\) is obtained by aggregating all incoming literals with \(\mathsf{op}_v\):
\begin{equation}
f_v(a,b)
=
\mathsf{op}_v\!\Bigl(\{x_u:\,u\in\mathcal{I}_v^+\}\cup\{\neg x_u:\,u\in\mathcal{I}_v^-\}\Bigr)
\end{equation}
Equivalently, \(f_v\) is the conjunction or disjunction of the literals contributed by its signed incoming edges. For nodes with no incoming regulators, we used neutral default values so that the update rule remains well-defined: \(f_v\equiv \mathrm{True}\) when \(\mathsf{op}_v=\mathrm{AND}\), and \(f_v\equiv \mathrm{False}\) when \(\mathsf{op}_v=\mathrm{OR}\).

\paragraph{Synchronous STG Construction.}
The synchronous STG is the directed graph \(G_F=(\mathcal{S},E)\) on the state space
\[
\mathcal{S}=\{(0,0),(0,1),(1,0),(1,1)\}
\]
with one deterministic outgoing edge from each state, namely
\[
x \to F(x)
\]
Thus, each admissible signed subgraph together with one operator pair defines exactly one synchronous Boolean system and one associated STG.
\paragraph{Asynchronous STG Construction.}
For the asynchronous analysis, we used a BoolNet-style one-node update rule. From a current state \(x=(a,b)\), exactly one node is selected and updated while all other nodes are kept fixed. Updating node \(A\) gives
\[
x \to \bigl(f_A(a,b),\,b\bigr),
\]
whereas updating node \(B\) gives
\[
x \to \bigl(a,\,f_B(a,b)\bigr).
\]
The state produced by this one-node update becomes the current state for any subsequent update; therefore asynchronous trajectories are sequential paths through the STG rather than parallel evaluations of all node updates from a fixed initial condition. The asynchronous STG is the union of all possible one-node transitions from each state. Self-loop transitions were omitted when a state had at least one non-self outgoing transition and retained only when no non-self transition was available, so that fixed points remain visible without allowing self-loops to obscure the nontrivial asynchronous dynamics.
\paragraph{Dynamical Feature Extraction.}
To quantitatively characterize the dynamics of each STG and prepare the data for subsequent statistical analyses or machine learning, we extracted numerical descriptors summarizing the topological and dynamical properties of the graphs. For synchronous STGs, attractors were fixed points or directed cycles of the deterministic update map. For asynchronous STGs, attractors were terminal strongly connected components, corresponding to trap components with no outgoing transition to states outside the component. The synchronous feature matrix is provided in \suppSfive{}, and the asynchronous feature matrix is provided in \suppSsix{}. In each case, the STG was represented as a feature vector $\mathbf{x}\in\mathbb{R}^p$, with the full set of feature definitions described in \suppStwo{}. Collecting the feature vectors for all STGs yields a feature matrix $\mathbf{X}\in\mathbb{R}^{N\times p}$, where $N$ is the number of STGs analyzed and $p$ is the number of extracted features.
 
\paragraph{Low-dimensional analysis of STG features extraction}
The numerical feature matrices $\mathbf{X}$ extracted from the synchronous and asynchronous STGs serve as the inputs for downstream statistical and manifold analyses. Prior to embedding, each matrix was systematically cleaned: missing values and infinite entries were handled, and non-informative features (zero variance) were removed. Full details of these preprocessing steps, scaling conventions, and reproducibility measures are provided in \suppSthree{}.

To obtain interpretable low-dimensional representations of the high-dimensional STG dynamics, Principal Component Analysis (PCA) was applied. PCA achieves two main objectives: (i) producing a compact two-dimensional embedding that preserves maximal variance for visualization, and (ii) guiding feature selection by identifying the features with the strongest contributions to principal components. Complementary nonlinear embeddings, including UMAP and t-SNE, were also computed to capture potential nonlinear structures in the data.

Reproducibility was promoted through deterministic preprocessing steps and fixed random seeds for stochastic algorithms, including K-means, t-SNE, and UMAP. The cleaned and robust-scaled STG feature matrices were used consistently across all dimensionality-reduction, clustering, and visualization analyses. This pipeline bridges STG-based feature extraction with manifold learning and statistical analysis, providing both visual and quantitative insight into the dynamical organization of the regulatory subgraphs under alternative update schemes.

\section{Results}
We begin by classifying all possible interactions between two Boolean nodes into seven structural connectivity classes, each representing a distinct topological skeleton (e.g., simple asymmetric regulation, mutual feedback, and variations with auto-regulatory self-loops; see Fig.~\ref{fig:two-node-classes-connected}). By assigning functional signs (activation or inhibition) to the directed edges and loops within these seven base structures, exactly 39 non-isomorphic signed digraphs were analytically identified. From these 39 graphs, all admissible two-node Boolean update rules consistent with their structural interaction patterns were enumerated, yielding 89 distinct Boolean rule sets. Each rule set defines a specific realization of the regulatory logic on a given graph. Results are reported in two layers: first, the deterministic synchronous STGs over the four possible Boolean states $(00,01,10,11)$; second, the corresponding asynchronous STGs generated by one-node-at-a-time updates on the same 89 networks.

Analysis of the 89 synchronous STGs revealed that, despite their distinct topologies and rules, they produce only 36 unique attractor configurations. This demonstrates substantial dynamical redundancy: multiple topologically or logically distinct models converge to identical or similar long-term behaviors in terms of attractor number, type, and basin organization. To capture finer distinctions among these synchronous STGs, 22 numerical dynamical descriptors were initially extracted for each graph. After removing two constant features, 20 retained nonconstant features were used for dimensionality reduction and clustering. These features were compiled into a synchronous feature matrix (\suppSfive{}), which was subsequently cleaned and standardized to support downstream analyses, including low-dimensional embeddings and comparative feature-based exploration. The asynchronous feature matrix and embeddings are reported separately in Section~\ref{subsec:async_stg_analysis} and \suppSsix{}. Full preprocessing details, including data cleaning and standardization methods, are provided in the Methods and \suppSthree{}.

\subsection{Exhaustive Instantiation of Two-Node Boolean Subgraphs}

Building upon the structural enumeration derived in Section 2.1 and the rule-counting framework of Section 2.2, we generated the complete set of admissible two-node BNs under the defined AND/OR rule catalogue. Each network corresponds to a non-isomorphic signed digraph with permissible loops, coupled with a valid Boolean rule assignment.

Figure~\ref{fig:bn-catalogue} presents the full catalogue of these BN structures. In total, 89 structurally and logically distinct BNs were instantiated. To uniquely identify each realization, we employ a systematic three-value indexing nomenclature (\textit{C}$x$--\textit{G}$y$--\textit{BN}$z$). In this classification scheme, $x$ denotes the parent structural connectivity class (1--7), $y$ identifies the specific graph number within that category, and $z$ designates the particular Boolean logical realization for that graph. Together, these three values establish the unique identity of each 89 instantiated network. Each diagram specifies both the signed interaction topology and its associated logical configuration.
\begin{figure}[htbp]
    \centering
\includegraphics[width= 1 \textwidth]{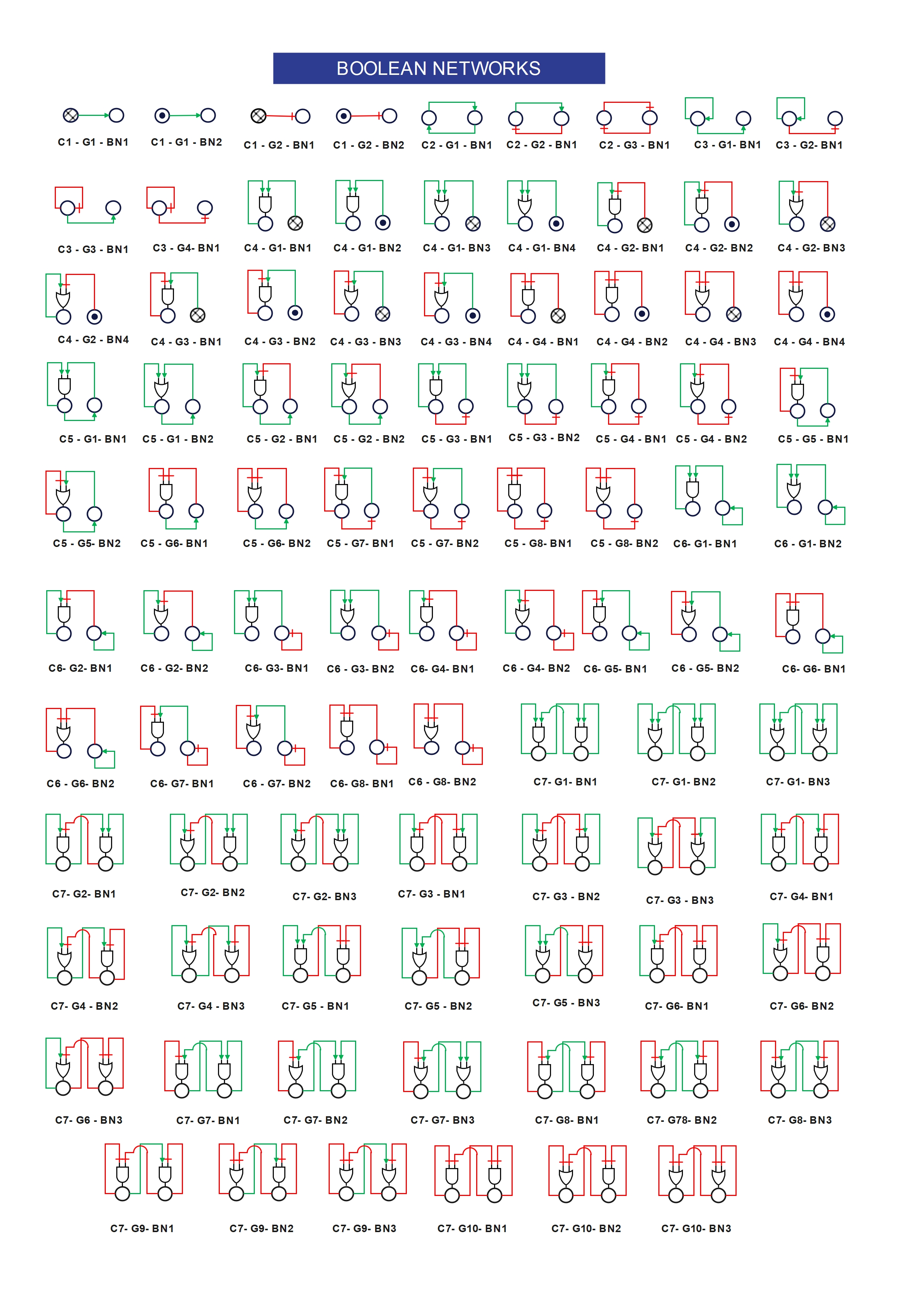}

\end{figure}
\clearpage
\begin{center}
    \captionof{figure}{\textbf{Complete catalogue of  Boolean Networks.} This figure presents the 89 structurally and logically distinct Boolean Networks constructed from the 39 non-isomorphic signed digraphs using the admissible AND/OR update-rule catalogue. Each panel represents a unique signed structural topology coupled with a specific Boolean-rule realization. Directed edges indicate regulatory interactions: green lines denote activation, whereas red lines denote inhibition. Self-loops indicate auto-regulation. Nodes receiving multiple regulatory inputs are shown with standard logic-gate symbols (AND/OR), indicating how the incoming signals are integrated. Empty circles denote Boolean state nodes, circles with a dot indicate state 0, and circles with a grid indicate state 1. Networks are systematically indexed as Cx--Gy--BNz, where C identifies the structural connectivity class (1--7), G identifies the signed graph topology within that class, and BN identifies the corresponding Boolean Network realization.}
    \label{fig:bn-catalogue}
\end{center}

For each BN, we constructed the synchronous State-Transition Graph (STG) evaluating the deterministic dynamical landscape over the state space 
$$ S = \{(0,0), (0,1), (1,0), (1,1)\}. $$

\begin{figure}[p]
    \centering
    \includegraphics[width=\textwidth,height=0.80\textheight,keepaspectratio]{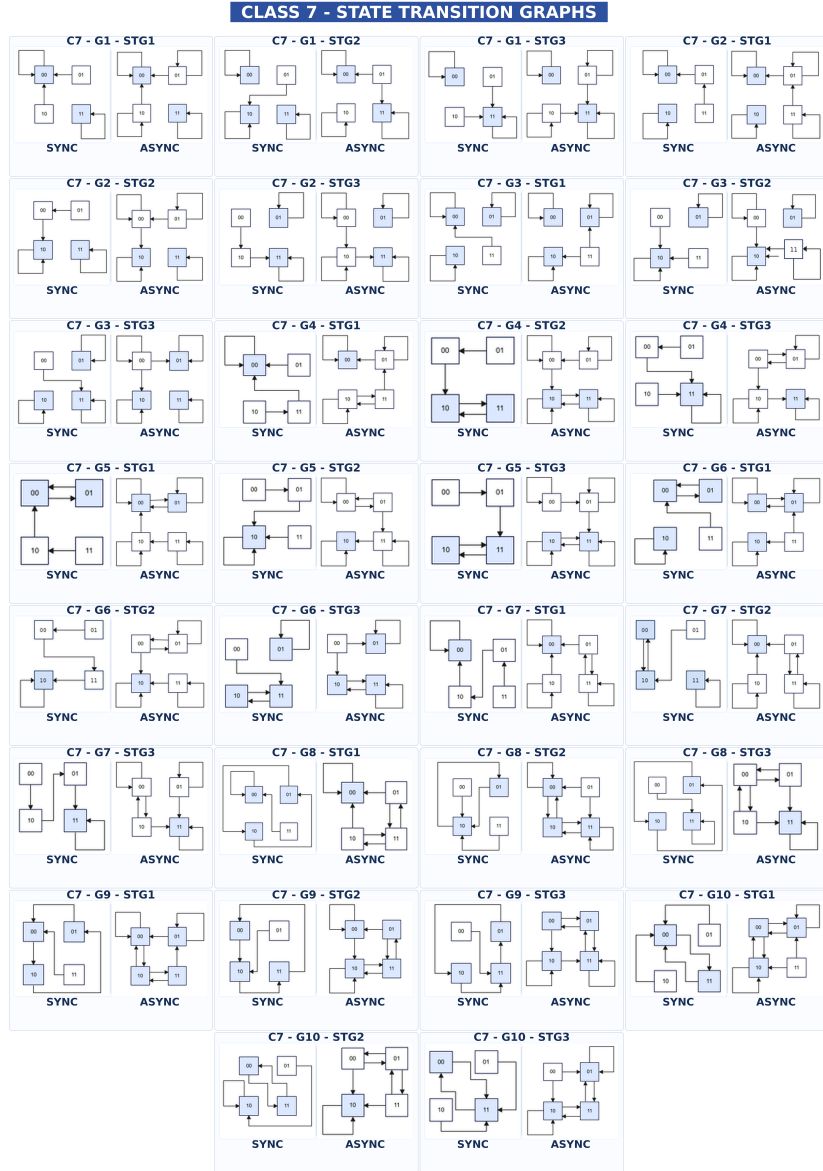}
    \caption{\textbf{Representative Class 7 state-transition graphs.} Because the complete set of 89 STGs is too large to display clearly in the main text, this figure shows Class 7 as a representative high-connectivity example with synchronous and asynchronous STGs shown side by side. Nodes represent the four global states \(00,01,10,11\); light blue nodes indicate attractor states and white nodes indicate transient states. The full class-wise visual catalogue is provided in \suppSone{}.}
    \label{fig:stg-catalogue}
\end{figure}

Figure~\ref{fig:stg-catalogue} shows Class 7 as a representative high-connectivity example of the STG catalogue, with synchronous and asynchronous state-transition graphs displayed side by side for the corresponding Boolean Network realizations. The full visual catalogue is provided in \suppSone{}. Each synchronous STG encapsulates the deterministic dynamical behavior of its corresponding BN under simultaneous updating of both nodes. Because the state space of a two-node Boolean network is strictly finite, encompassing $2^N = 4$ possible global states ($S \in \{00, 01, 10, 11\}$), the deterministic, synchronous update rules guarantee that the system will eventually converge to a repeating long-term behavior known as an attractor. By mapping all possible state transitions, these directed graphs explicitly delineate the network's dynamical components. Specifically, the STGs capture stable fixed-point attractors, defined as equilibrium states that transition directly into themselves and satisfy the mathematical condition $S(t+1) = S(t)$. Furthermore, they depict limit cycles (cyclic attractors), which emerge as oscillatory loops where a sequence of distinct states transition into one another, governed by $S(t) \to S(t+1) \to \dots \to S(t)$ for a period of $k > 1$. Preceding these attractors are transient trajectories, representing sequences of non-repeating states that the network passes through before ultimately settling. Together, an attractor and its incoming transient pathways constitute a basin of attraction, defined as the complete subset of initial states that eventually funnel into that specific stable outcome. The asynchronous counterpart is not deterministic in the same sense, because each state can branch according to whether node \(A\) or node \(B\) is selected for update; those asynchronous dynamics are therefore summarized separately in Section~\ref{subsec:async_stg_analysis}.

Together, these visual catalogues establish an exhaustive structural-to-dynamical
mapping, laying the formal foundation for the subsequent statistical and clustering
analyses. To ensure transparency and reproducibility, the complete class-wise
synchronous and asynchronous visual enumeration is provided in \suppSone{}. This
supplementary catalogue organizes all 89 Boolean Network realizations by structural
class and displays, for each case, the signed regulatory topology, Boolean update
functions, and corresponding STGs under the update schemes considered here. Thus,
whereas Fig.~\ref{fig:bn-catalogue} and Fig.~\ref{fig:stg-catalogue} provide compact
visual overviews, \suppSone{} provides the full visual class-wise catalogue. The
corresponding asynchronous transition and feature data are provided in \suppSsix{}.

\subsection{Visualization of Synchronous Clusters in Feature Space}
\begin{figure}[htbp]
\centering
\includegraphics[width=1\textwidth]{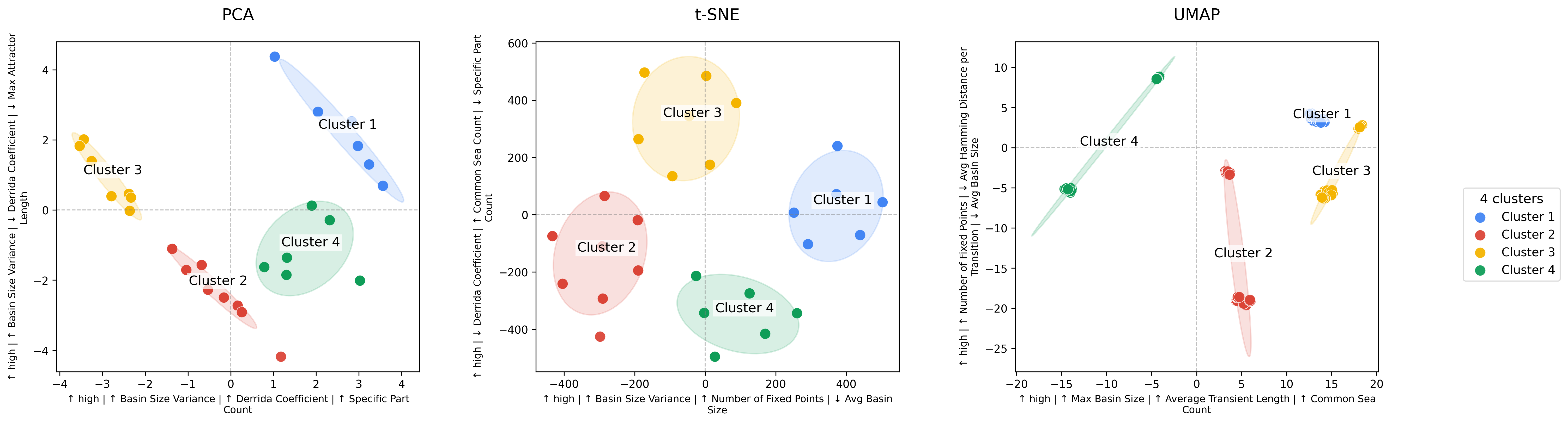} 
\caption{\textbf{Dimensionality reduction and K-means clustering ($K=4$) of the 20-feature synchronous STG dataset.} PCA, t-SNE, and UMAP projections were computed from the 20 retained robust-scaled synchronous STG descriptors. Axis labels indicate the top three features or correlations driving high/low values along each dimension.}
\label{fig:clusters-all-features}
\end{figure}

To determine whether two-node Boolean Networks with different structures and
rules occupy distinct regions of dynamical-feature space, we projected the
20-feature dataset into two dimensions using PCA, t-SNE, and UMAP and overlaid
the four K-means clusters. Figure~\ref{fig:clusters-all-features} illustrates how
these clusters are distributed across the resulting reduced feature spaces. Each embedding is annotated with the dominant feature gradients along its axes, allowing us to interpret how increases or decreases in individual features correspond to the spatial organization of the clusters.

In the PCA projection, the horizontal axis reflects a transition from high Basin Size Variance and Derrida Coefficient on the left to high Specific Part Count on the right. Basin Size Variance measures how unevenly initial states are distributed among attractor basins, whereas the Derrida Coefficient quantifies the average sensitivity of the dynamics to small perturbations. The vertical axis is driven primarily by Max Attractor Length, which records the longest fixed-point or cyclic attractor observed for a given network. Under these gradients, the clusters occupy distinct regions of the synchronous feature space, separating stable fixed-point-dominated networks from oscillatory and composite attractor regimes.

The t-SNE and UMAP projections provide complementary nonlinear views of the same 20-feature matrix. Because these embeddings do not have linear principal-component axes, their axis labels are interpreted as correlation-based feature guides rather than direct mechanistic axes. Nevertheless, both nonlinear embeddings preserve the same broad organization seen in PCA: networks with similar attractor structure and basin organization remain grouped together, while stable, oscillatory, complex composite, and hybrid regimes occupy separate regions of feature space.

Across all three visualizations, the four clusters maintain consistent separation along dominant feature gradients. This agreement indicates that the cluster structure is robust and reflects meaningful regimes in the underlying Boolean network feature space, rather than artifacts of a particular dimensionality-reduction method.

\subsection{Synchronous Attractors}

To systematically characterize the deterministic dynamical landscape of the enumerated subgraphs, we classified each synchronous STG according to its attractor structure. For a deterministic Boolean system with two nodes, the global state space contains four states,
\[
S=\{00,01,10,11\}.
\]
At the level of arbitrary deterministic maps on this four-state space, 51 attractor configurations are theoretically possible, including fixed-point sets, cyclic attractors, and mixed fixed/cyclic configurations. However, the present enumeration does not span all possible deterministic maps \(F:S\to S\). Instead, it is restricted to connected two-node signed regulatory graphs whose node update functions are generated by signed inputs combined through AND/OR aggregation, with fixed neutral defaults for nodes without incoming regulators. Consequently, only a subset of all possible four-state transition maps can be produced.

Within this restricted graph-rule catalogue, the 89 enumerated synchronous STGs realize 36 of the 51 theoretically possible deterministic attractor configurations. The remaining 15 configurations are therefore not impossible for Boolean Networks in general; rather, they are not reachable under the specific structural and logical constraints imposed in this study. In particular, some unrealized configurations would require transition maps that cannot be represented using only signed-literal AND/OR rules on connected two-node subgraphs. These configurations could become realizable if more general Boolean functions, such as XOR/XNOR-like dependencies or arbitrary truth-table rules, different default conventions, larger subgraphs, or alternative update schemes were allowed. The complete list of the 51 theoretical configurations, including the 36 realized synchronous and 15 unrealized deterministic cases, is provided in \suppSfour{}.\\
The Boolean networks in our dataset exhibit four characteristic attractor types: 
\begin{itemize}
    \item \(Attractor Type_1\) – Complex Composite Regime
Enriched in mixed and higher-order composite attractors, indicative of distributed and heterogeneous dynamics.
    \item \(Attractor Type_2\) – Pure Oscillatory Regime
Defined by exclusively cyclic attractors, representing uniform periodic behavior.
 \item \(Attractor Type_3\) – Stable Fixed-point Regime
Dominated by steady states, reflecting strong convergence and dynamical stability.
 \item \(Attractor Type_4\) – Hybrid Multimodal Regime
Characterized by coexisting fixed and composite attractors, capturing intermediate dynamical complexity.
\end{itemize}

Clustering of the STG partitioned the Boolean networks into four groups with clearly distinct attractor landscapes (Table \ref{tab:attractor_cluster_mapping}). Cluster 1 (\(n = 21\) networks) is dominated by heterogeneous and composite structures, including 14 complex composite attractor configurations and 7 hybrid multimodal configurations, with no pure oscillatory or stable fixed-point-only cases, indicating a regime of structurally complex and distributed dynamics. In contrast, Cluster 2 (\(n = 24\)) consists exclusively of pure oscillatory configurations, representing a periodic regime with no steady-state-only or mixed configurations. Cluster 3 (\(n = 26\)) is strongly enriched in fixed points, with 22 stable fixed-point configurations and 4 pure oscillatory configurations, corresponding to a largely stable regime dominated by steady-state convergence. Cluster 4 (\(n = 18\)) exhibits a heterogeneous composition, including 9 complex composite configurations and 9 hybrid multimodal configurations, while lacking pure oscillatory and stable fixed-point-only cases, suggesting a hybrid regime in which composite and mixed dynamical organization coexist. Collectively, these results reveal a clear stratification of dynamical regimes across clusters, spanning purely oscillatory (Cluster 2), strongly stable (Cluster 3), and increasingly complex, multi-attractor organizations (Clusters 1 and 4), highlighting the ability of the clustering approach to resolve functionally distinct dynamical classes within Boolean network state spaces.
\begin{table}[h!]
\centering
\caption{Distribution of synchronous attractor types across the four clusters.}
\label{tab:attractor_cluster_mapping}

\begin{tabular}{lccccc}
\toprule
\textbf{Clusters} & 
\makecell{Stable \\ Fixed-point} & 
\makecell{Pure \\ Oscillatory} & 
\makecell{Complex \\Composite } & 
\makecell{Hybrid \\ Multimodal} & 
\textbf{Total} \\
\midrule
\(Cluster_1\) & 0 & 0 & 14 & 7 & 21 \\
\(Cluster_2\) & 0 & 24 & 0 & 0 & 24 \\
\(Cluster_3\) & 22 & 4 & 0 & 0 & 26 \\
\(Cluster_4\) & 0 & 0 & 9 & 9 & 18 \\
\midrule
\textbf{Total} & 22 & 28 & 23 & 16 & 89 \\
\bottomrule
\end{tabular}

\end{table}

The comparison between synchronous and asynchronous updating is particularly informative in this reduced setting because it allows us to distinguish dynamical behaviors that are robust across update schemes from those that depend on a particular update schedule. In particular, it can reveal synchronous limit cycles that disappear under asynchronous updating, thereby identifying behaviors that may be artifacts of the assumption of simultaneous updating. Conversely, dynamical behaviors that persist across both update schemes may represent more robust properties of the underlying network motif. This comparison therefore provides a systematic way to assess the extent to which the observed dynamics depend on the assumed update scheme and to identify which dynamical features are preserved when the strict synchrony assumption is relaxed.
\subsection{Asynchronous State-Transition Graph Analysis}
\label{subsec:async_stg_analysis}

To complement the deterministic synchronous analysis, we next evaluated the same 89 Boolean Network realizations under an asynchronous update scheme. In this setting, each transition updates exactly one node at a time. Thus, from a current state \(x=(a,b)\), the asynchronous STG contains the possible one-node transitions
\[
x \to \bigl(f_A(a,b),b\bigr),
\qquad
x \to \bigl(a,f_B(a,b)\bigr),
\]
corresponding respectively to updating node \(A\) only or node \(B\) only. Self-loops were removed whenever a state had at least one non-self outgoing transition, and were retained only when no non-self transition was available. This convention preserves fixed points while preventing uninformative self-loops from dominating the asynchronous transition structure.

The asynchronous analysis was performed on the same 39 connected signed graph classes and 89 Boolean rule realizations used in the synchronous catalogue. As in the synchronous analysis, each network was represented by a quantitative STG-derived feature vector and embedded using PCA, t-SNE, and UMAP followed by \(K=4\) K-means clustering. In contrast to the synchronous case, asynchronous STGs are generally nondeterministic because a state may have more than one possible successor depending on which node is selected for update. Consequently, attractors were identified as terminal strongly connected components of the asynchronous STG, corresponding to trap components with no outgoing transitions to states outside the component.

Across the 89 asynchronous STGs, 27 distinct attractor configurations were observed. Most networks had a single terminal attractor (\(57/89\)), while 25 networks had two attractors and 7 networks had three attractors. Fixed-point attractors were present in 61 networks, whereas 35 networks contained at least one non-singleton terminal component, interpreted here as an asynchronous limit-cycle or trap component. The maximum attractor length was 1 for 54 networks, 2 for 18 networks, 3 for 3 networks, and 4 for 14 networks. Thus, asynchronous updating preserved substantial dynamical redundancy across the catalogue, while also producing broader terminal trap components than those observed under deterministic synchronous updating.

For feature-space analysis, 41 numerical asynchronous STG descriptors were computed. Two constant descriptors were removed, leaving 39 retained nonconstant features for robust scaling, dimensionality reduction, and clustering. The resulting asynchronous feature matrix and low-dimensional embeddings are provided in \suppSsix{}. The four-cluster solution separated the networks into groups of sizes \(25\), \(14\), \(21\), and \(29\) for Cluster 1--Cluster 4, respectively. Cluster 1 was enriched for multi-attractor fixed-point organization. Cluster 2 corresponded to fully non-fixed cyclic or trap-component behavior. Cluster 3 represented a mixed asynchronous regime, combining cyclic/trap-dominated networks with mixed fixed-point plus trap networks. Cluster 4 was dominated by single fixed-point attractors.

As summarized in Table~\ref{tab:async_attractor_cluster_mapping}, the asynchronous feature-space clustering separated the 89 Boolean Network realizations into four dynamical regimes. Clusters 1, 2, and 4 were homogeneous under this attractor-type classification, whereas Cluster 3 contained cyclic/trap-only networks and mixed fixed-point plus cyclic/trap networks.

\begin{table}[h!]
\centering
\caption{Distribution of asynchronous attractor types across the four clusters.}
\label{tab:async_attractor_cluster_mapping}
\begin{tabular}{lccccc}
\toprule
\textbf{Clusters} &
\makecell{\textbf{Single}\\\textbf{fixed point}} &
\makecell{\textbf{Multi-attractor}\\\textbf{fixed-point}} &
\makecell{\textbf{Cyclic / trap}\\\textbf{dominated}} &
\makecell{\textbf{Mixed}\\\textbf{fixed + trap}} &
\textbf{Total} \\
\midrule
\(Cluster_1\) & 0  & 25 & 0  & 0 & 25 \\
\(Cluster_2\) & 0  & 0  & 14 & 0 & 14 \\
\(Cluster_3\) & 0  & 0  & 14 & 7 & 21 \\
\(Cluster_4\) & 29 & 0  & 0  & 0 & 29 \\
\midrule
\textbf{Total} & 29 & 25 & 28 & 7 & 89 \\
\bottomrule
\end{tabular}
\end{table}

\begin{figure}[htbp]
\centering
\includegraphics[width=1\textwidth]{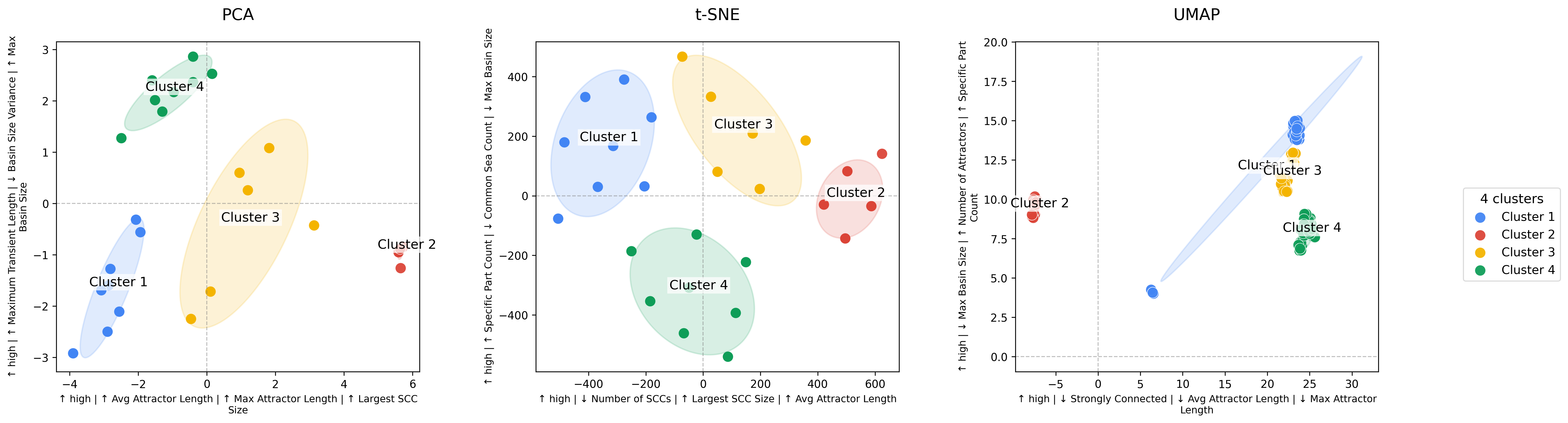}
\caption{\textbf{Dimensionality reduction and K-means clustering (\(K=4\)) of the asynchronous STG feature matrix.} PCA, t-SNE, and UMAP projections were computed from the 39 retained robust-scaled asynchronous STG descriptors. Points represent the same 89 Boolean Network realizations analyzed under asynchronous updating. Colors denote K-means clusters using the same Google-color template as the synchronous feature-space visualization. Axis labels indicate the dominant feature gradients or correlation-based feature directions for each embedding.}
\label{fig:async-clusters-all-features}
\end{figure}

\clearpage

\subsection{Comparison of Synchronous and Asynchronous Dynamics}

Because the synchronous and asynchronous analyses were performed on the same 89 graph-rule realizations, their attractor classifications can be compared network by network. The comparison shows that fixed-point organization is highly robust to the update scheme. All 22 networks with a single synchronous fixed-point attractor remained single fixed-point systems under asynchronous updating, and all 23 networks with multiple synchronous fixed-point attractors remained multi-attractor fixed-point systems. Similarly, the 28 networks classified as purely oscillatory under synchronous updating retained non-singleton terminal components under asynchronous updating, although the asynchronous attractors are interpreted as terminal trap components rather than deterministic cycles.

The main update-dependent behavior occurred in the synchronous hybrid multimodal class. Of the 16 networks containing both fixed-point and cyclic attractors under synchronous updating, only 7 retained a mixed fixed-point plus trap structure under asynchronous updating. The remaining 9 lost their non-singleton terminal component: 2 became multi-attractor fixed-point systems and 7 became single fixed-point systems. Thus, the strict assumption of simultaneous updating most strongly affects mixed regimes in which fixed-point and cyclic behaviors coexist. In these cases, a synchronous limit cycle can represent a schedule-dependent behavior that is not preserved when the system is allowed to update one node at a time.

This comparison clarifies the role of the asynchronous analysis. It is not merely a second visualization of the same catalogue, but a test of update-scheme robustness. Behaviors preserved under both update schemes represent stronger candidates for intrinsic motif-level dynamics, whereas behaviors that change under asynchronous updating should be interpreted as update-dependent. This distinction is especially important when connecting the two-node catalogue to larger Boolean models, where biological components rarely update in perfect synchrony.

\subsection{BORNA: Interactive exploration of two-node Boolean network dynamics}
\begin{figure}[htbp]
\centering
\includegraphics[width=1\textwidth]{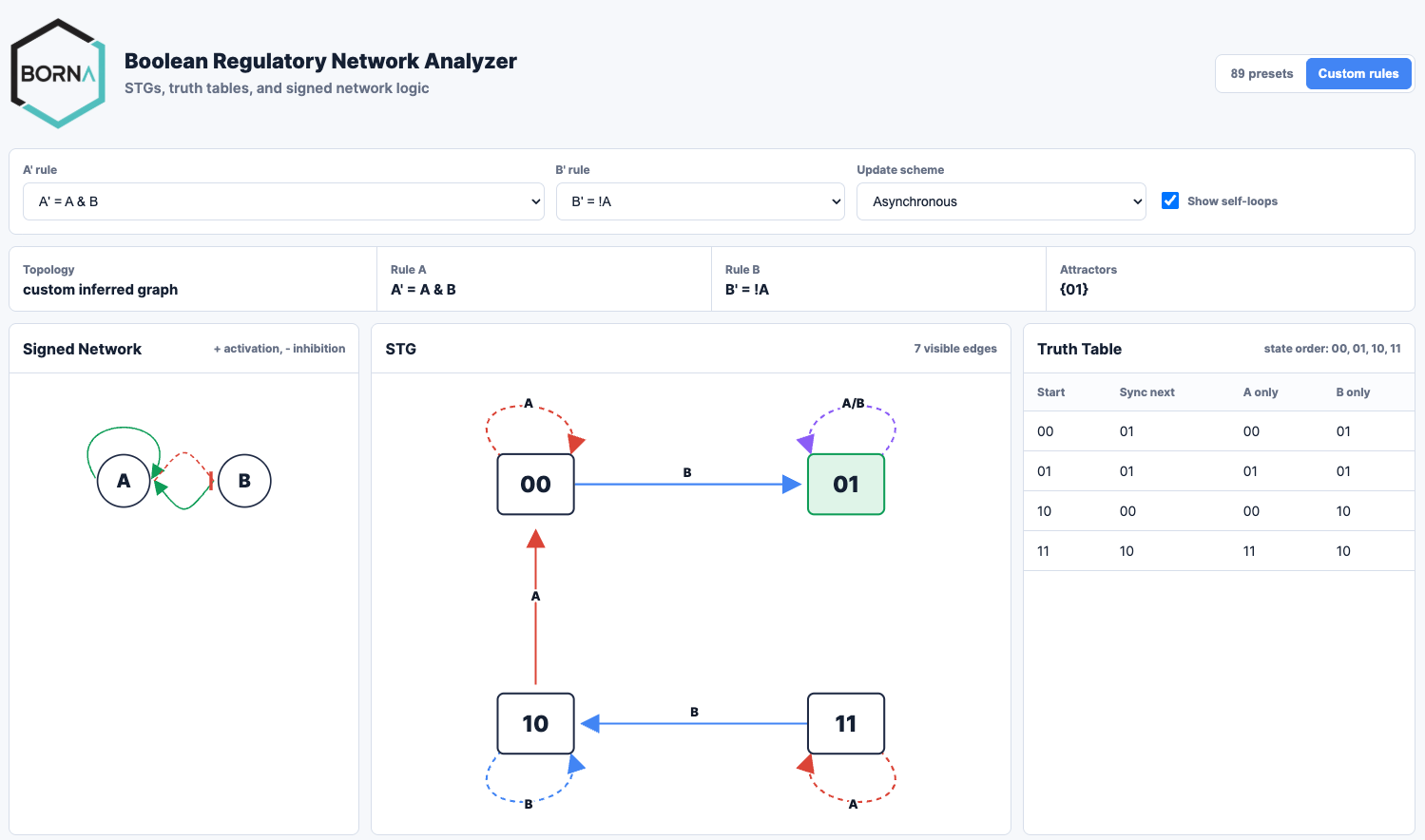}
\caption{\textbf{Screenshot of the BORNA web application.} BORNA provides interactive exploration of two-node signed Boolean network structures, their corresponding state-transition graphs (STGs), dynamical classifications, and simulations under both synchronous and asynchronous updating. The application integrates structural and dynamical information within a unified interface, enabling users to examine how network topology and update schemes influence emergent dynamical behavior.}
\label{fig:borna}
\end{figure}

To facilitate exploration of the enumerated networks and their dynamical behaviors, we developed BORNA (BOolean network two-node dynamics via state-tRansition graph aNAlysis, https://jafarilab.github.io/BORNA/), an interactive web application that provides access to the complete catalogue of two-node signed Boolean networks analyzed in this study (Fig.~\ref{fig:borna}). The application allows users to explore network structures and their corresponding state-transition graphs, inspect the associated dynamical classifications, and simulate the networks under both synchronous and asynchronous updating. By integrating structural information with STG-based dynamical analysis in an interactive environment, BORNA provides a practical interface for exploring the relationships between network topology, update scheme, and emergent dynamics without requiring users to reproduce the computational pipeline independently.

\label{subsec:borna}

\clearpage
\section{Discussion}

This study provides a systematic catalogue of the dynamical behaviors generated by two-node signed Boolean networks under the structural and logical constraints considered here. By combining exhaustive enumeration with state-transition graph (STG)-based analysis, we found that 89 graph-rule realizations give rise to 36 distinct synchronous attractor configurations, demonstrating that substantially different network structures can converge to a more limited repertoire of dynamical behaviors. The resulting dynamical classes provide a compact representation of this structural diversity and reveal recurring regimes including stable, oscillatory, and mixed attractor organizations.

The comparison between synchronous and asynchronous updating provides an additional perspective on the robustness of these dynamical behaviors. Applying asynchronous updating to the same 89 realizations produced 27 distinct attractor configurations and four broader dynamical regimes. This comparison allows us to distinguish behaviors that are preserved across update schemes from those that depend on synchronous updating. In particular, synchronous limit cycles that disappear under asynchronous updating may represent update-dependent behaviors. In contrast, features preserved under both schemes may reflect more robust properties of the underlying regulatory motifs. Thus, the asynchronous analysis is not simply an extension of the catalogue but provides a way to assess the sensitivity of motif dynamics to assumptions about temporal updating.

The present catalogue does not yet establish a direct reduction method for large Boolean networks. In larger systems, the behavior of a pair of nodes can depend on external regulators, intermediate nodes, logical functions, and feedback from the surrounding network. Nevertheless, the catalogue suggests a testable strategy for connecting local regulatory structure to dynamical behavior. For a selected pair of nodes (A) and (B), an effective local regulatory subnetwork could be constructed by considering directed paths from (A) to (B), paths from (B) to (A), and feedback cycles involving either node. These relationships could then be summarized in terms of effective regulatory direction, sign, reciprocal interactions, and self-regulation, and subsequently compared with the dynamical classes in the present catalogue. Path-based approaches for quantifying signed regulatory influence provide a potential methodological starting point for constructing such effective representations \cite{campbell2011network}. Whether such an effective representation can reliably predict local dynamics remains an open question and will require validation against full-network simulations and established reduction methods.

To facilitate exploration and reproducibility, we also developed the \textbf{BORNA} web application, which provides interactive access to the complete catalogue of two-node networks and their corresponding STGs. BORNA allows users to explore network structures, dynamical classifications, and both synchronous and asynchronous simulations without requiring independent implementation of the analysis pipeline. The application is intended to make the catalogue accessible as a practical resource for researchers interested in Boolean network dynamics and to support further investigation of the proposed connections between local dynamical motifs and larger regulatory systems.

The framework presented here provides a foundation for extending the analysis to three-node and larger subnetworks, as well as to alternative update schemes and more general Boolean rules. Such extensions will help determine whether the observed compression of structural diversity into a smaller repertoire of dynamical behaviors persists at higher network orders and whether the resulting catalogue can ultimately contribute to the analysis and reduction of larger Boolean regulatory networks.

\section*{Acknowledgments}

We sincerely thank Réka Albert for her careful review of the manuscript, thoughtful recommendations, and valuable guidance throughout the revision process. We greatly appreciate her time, expertise, and generous support in helping us improve this work.

\section*{Data and Code Availability}

The complete catalogue of two-node signed Boolean network realizations and the data underlying the synchronous and asynchronous state-transition graph (STG) analyses are available at the BORNA GitHub repository: \url{https://github.com/jafarilab/BORNA}. The repository also contains the source code for network enumeration, Boolean rule generation, synchronous and asynchronous STG analysis, attractor identification, feature extraction, and dynamical classification.

In addition, the repository includes the source code for the BORNA web application, \textbf{BOolean network two-node dynamics via state-tRansition graph aNAlysis}, which provides an interactive interface for exploring the network catalogue and its dynamical behaviors. The application is freely accessible at \url{https://jafarilab.github.io/BORNA/}.

\section*{Supplementary Material}
The supplementary material is organized as independently compiled PDF files. Each supplementary source file is kept as a separate \texttt{.tex} document, and the corresponding PDF should be uploaded alongside the main manuscript during submission.

\begin{itemize}
    \item \suppSone{}: Visual catalogue of Boolean Network classes and their corresponding synchronous and asynchronous state-transition graphs.
    \item \suppStwo{}: Definitions of the dynamical features extracted from the state-transition graphs.
    \item \suppSthree{}: Feature-matrix preprocessing, dimensionality reduction, clustering, and reproducibility pipeline.
    \item \suppSfour{}: Theoretical and realized attractor configurations for the two-node Boolean Network state space.
    \item \suppSfive{}: Summary of the raw and processed synchronous STG feature matrices.
    \item \suppSsix{}: Asynchronous STG feature matrix, cluster assignments, embedding coordinates, and PCA/t-SNE/UMAP visualization.
\end{itemize}

The machine-readable data matrices associated with \suppSfive{} are provided as separate CSV files: \suppSfiveRaw{}, \suppSfiveRetained{}, \suppSfiveScaled{}, \suppSfiveClusters{}, and \suppSfiveEmbeddings{}.

The machine-readable data matrices associated with \suppSsix{} are provided as separate CSV files:
\begin{itemize}
    \item \suppSsixRaw{}
    \item \suppSsixScaled{}
    \item \suppSsixClusters{}
    \item \suppSsixEmbeddings{}
\end{itemize}

\bibliographystyle{unsrt}
\bibliography{references}

\end{document}